\documentclass[aps,superscriptaddress,twocolumn]{revtex4-2}
\usepackage[utf8]{inputenc}  
\usepackage[T1]{fontenc}
\usepackage[main=english]{babel}  
\usepackage[a4paper,vmargin=2.2cm,hmargin=2.0cm]{geometry}

\usepackage{longtable}

\usepackage{amsmath,amssymb,amsthm,amsfonts}  
\usepackage{mathrsfs}    
\usepackage{siunitx}     
\usepackage{braket}
\usepackage{cancel}
\usepackage{verbatim}     

\usepackage{graphicx}    
\usepackage{float}       
\usepackage{hyperref}
\usepackage{tabularx}
\usepackage{multirow}
\usepackage{colortbl}
\usepackage{subfigure}
\usepackage{enumitem}
\setdescription{labelsep=\linewidth,leftmargin=0pt}
\usepackage{leftindex}
\usepackage{nicefrac}

\usepackage[dvipsnames]{xcolor}
\usepackage{soul}

\newcommand{\LSrm}[1]{\textcolor{red}{\st{#1}}}

\newcommand{\vc}[1]{\mathbf{#1}}

\begin{document}

\title{First-principle investigation of the electronic structure and optical properties of graphene/boron nitride lateral heterostructures}

\author{Elisa Serrano Richaud}
\affiliation{Universit\'e Paris-Saclay, ONERA, CNRS, Laboratoire d'\'etude des microstructures (LEM), 92322 Ch\^atillon, France}

\author{Sylvain Latil}
\affiliation{Universit\'e Paris-Saclay, CEA, CNRS, SPEC, 91191 Gif-sur-Yvette, France}

\author{Lorenzo Sponza}
\affiliation{Universit\'e Paris-Saclay, ONERA, CNRS, Laboratoire d'\'etude des microstructures (LEM), 92322 Ch\^atillon, France}
\address{European Theoretical Spectroscopy Facility (ETSF), B-4000 Sart Tilman, Li\`ege, Belgium}

\begin{abstract}
We investigate the electronic and optical properties of lateral heterostructures made of alternated armchair ribbons of graphene and hexagonal boron nitride. 
It is known that the gapwidth of these heterostructures can be classified into three families depending on the width of the graphene part.
Here, by employing ab initio methods (standard and time-dependent density functional theory and GW), we demonstrate that such classification still holds for other electronic states close to the gap. 
We show that they display trends substantially different from those known for the gapwidth and originate family-specific features in the screening properties and optical absorption spectra (peak energy and intensity).
In addition, our use of a tight binding model originally introduced for isolated nanoribbons allows us to discuss some crucial heterostructure's properties in view of those of its isolated building blocks, including charge redistribution at the edges, gap hierarchy inversion, and specific optical selection rules.
By bridging the electronic structure to optical absorption spectra in a comprehensive set of systems, this study sets the stage for more refined investigations on the absorption properties of graphene/boron nitride lateral heterostructures.

\end{abstract}


\maketitle



Vertical heterostructures (or van der Waals heterostructures) are the most common kind of 2D heterostructures~\cite{Geim2013}. In these vertical assemblies, sheets of 2D materials are stacked one on top of the other.
At variance, in lateral heterostructures (LHS), the 2D materials are grown sequentially side-by-side which gives rise to a 1D interface with unique properties~\cite{Wang2019,Avalos-Ovando2019}.
The 
composition and shape of this monodimensional interface has a crucial role in the LHS properties. 
For instance, it has an influence on the exciton transfer across graphene/transition metal dichalcogenide interfaces~\cite{Razaghi_2023,Liu_2019,Fu_2020,Wu_2021,Chen_2019} or on electron-hole splitting in transition metal dichalcogenide based heterostructures~\cite{Li_2017,Ye_2019}.
Concerning the shape of the contact edges, the most common in honeycomb 2D materials are zigzag and armchair interfaces, although also more complex junctions can be designed~\cite{Li_2013,Nguyen_2019,Wang_2021}.
Changing this parameter leads to LHS with different properties, as it is the case in graphene/hexagonal boron nitride (Gr/hBN) junctions: 
while zigzag ones may present selective half-metallicity~\cite{Pruneda_2010,Dutta_2009,Pizzochero_2024,Leon_2019,Padilha_2013}, or being gapped~\cite{Bernardi_2012b}, armchair heterostructures are always semiconducting with a tunable gap~\cite{Bernardi_2012b,Seol_2011,Guan_2021,Nascimiento_2019}.
These features make armchair Gr/hBN LHS particularly interesting for possible applications as resonant tunneling diodes~\cite{Sanaeepur_2020,Ghassemian_2024,Yazdanpanah_2018}, field effect transitors~\cite{Dong_2012,Mehri_2017,Levendorf_2012,Jamaati_2017, Liu_2013,Wang_2021,Saraswat_2021}, or photodetectors~\cite{Bernardi_2012,Osella_2012,Zarei_2018}.
In addition, the fact that charge and energy transfers across the interface occur on the same plane as the rest of the heterostructure enables unique device architectures and geometries that can not be realised in vertical stacks.


Their synthesis has achieved nowadays a high level of control over the size and shape of material domains.
At its early stages, the so-called one-step growth was performed via chemical vapor deposition (CVD) and produced hybrid carbon-boron-nitrogen (CBN) domains with random size and shape~\cite{Ci_2010,Gong_2014}.
In the span of few years the community reached the point where precise domain shape can be achieved by a two-step growing process~\cite{Wang_2021,Saraswat_2021,Stehle_2017,Liu_2014,Levendorf_2012,Liu_2013}.
This consists in growing CVD graphene exploiting the edges of a previously grown hBN flake  as nucleation sites (or vice-versa).
A possible patterning step can be implemented between the two growths to sharpen the edges of the seed.
A similar, two-step growth has been implemented also in molecular beam epitaxy~\cite{Li_2016,Thomas_2020}.

\begin{figure}[b]
    \centering
    \includegraphics[width=\linewidth]{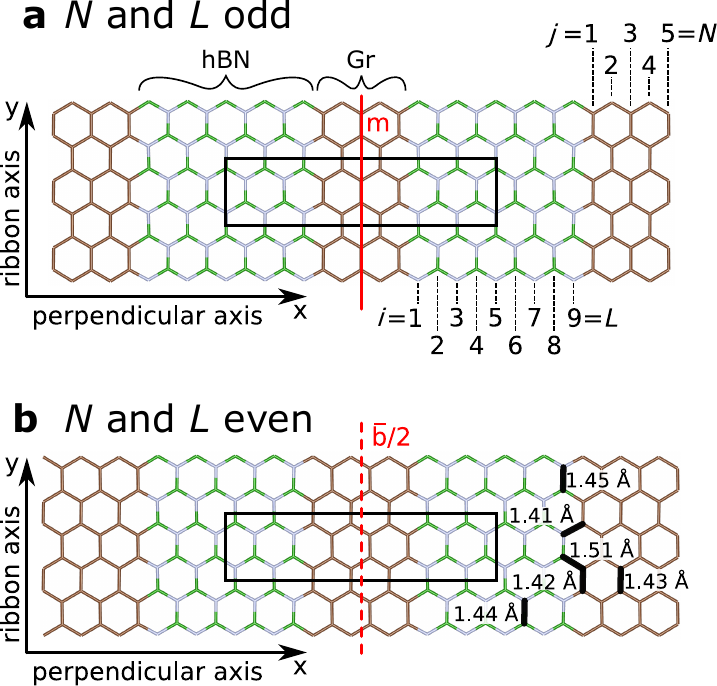}
    \caption{\textbf{a}: AGBN composed of graphene nanoribbons  with $N$=5 (brown sticks) and hBN nanoribbons with $L$=9 (green-and-grey sticks). The mirror symmetry plane (m) is reported with a red solid line. Row indexes are also reported as $1 \le j \le $$N$ in the graphene ribbon, and $1 \le i \le $$L$ in the BN ribbon.
    {\bf b}: AGBN with $N$=6 and $L$=9. A dashed red line indicates the glide symmetry plane ($\bar{b}$/2). Notable interatomic distances after relaxation are reported.
    In both panels, the simulation cell is reported as a solid black rectangle as well as the Cartesian reference indicating the perpendicular $x$ and the ribbon (parallel) $y$ axes.}
    \label{fig:ag5bn9_structure}
\end{figure}

\begin{figure*}
    \centering
    \includegraphics[width=\linewidth]{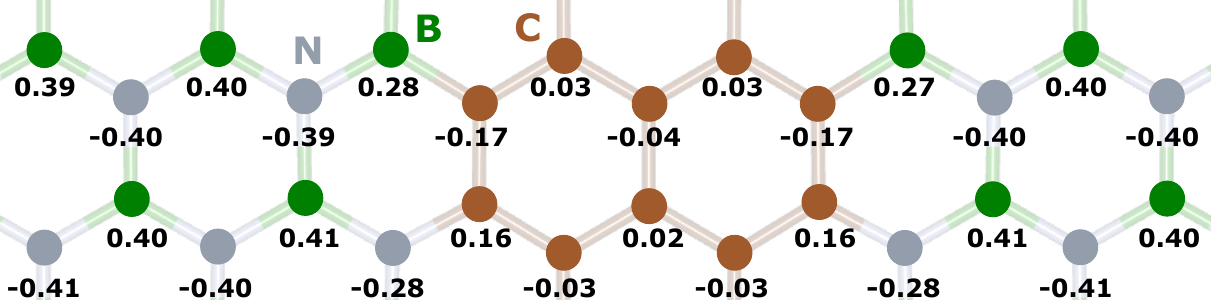}
    \caption{Voronoi deformation density analysis reporting the integrated charge of each atom (green for B, brown for C and grey for N). Excess and missing charges with respect to the nominal one are indicated below each atom.}
    \label{fig:Voronoi_analysis}
\end{figure*}

Many previous works on electronic and optical properties of Gr/hBN LHS focused on a specific kind of arrangement formed of an alternated sequence of armchair nanoribbons of Gr and hBN with different width (AGBN).
Even though much more complex design can be achieved experimentally, it is worth focusing on this simple arrangement since it is an ideal platform to study how confinement and the Gr/hBN interface impact the optoelectronic properties of more generic Gr/hBN lateral heterostructures.

The fundamental gap of AGBNs behaves quite similarly to that of armchair graphene nanoribbons. 
For example, the gapwidth versus graphene width displays the same oscillating behaviour as in isolated Gr armchair nanoribbons and can be decomposed in three separate trends, called families~\cite{Guan_2021,Seol_2011}.
Apart from this, other properties, as the specific role played by Gr and hBN parts or the changes on the optical spectra due to the Gr/hBN interface, were not studied in deep though they are crucial to fully understand AGBNs optical properties.
In this paper we address these questions by coupling first principle calculations in the framework of density functional theory (DFT) and G$_0$W$_0$ approximation with tight-binding modeling~\cite{SerranoRichaud2024}.

More precisely, in this article we want to achieve a thorough characterisation of the absorption onset of AGBNs, trace the origin of the main peaks back to the underlying electronic structure and derive models to gain insight and permit further investigations.
To this aim, it is necessary to unveil how these systems differ from the corresponding isolated nanoribbons, how far the classification into families can be pushed and to what extent it can help predicting the properties of the heterostructure.
In future works, that are currently under drafting, we will focus on excitons in ideal systems and in defective heterostructures with the ultimate goal of tracing the connections linking ground-state properties of ideal systems to accurate excitations in structures that may be relevant experimentally.

The paper is divided as follows:
in the first part, we focus on the structure of this class of systems. 
We detail their symmetries,  our treatment of the interface strain and we discuss the charge distribution. 
In the second part, we report on the electronic structure and scrutinize the role of each material. 
We also calculate quasiparticle corrections and we discuss the insight they give on screening properties.
In the third part, we calculate the optical selection rules of AGBNs and their absorption spectra in the independent particle and random phase approximations.
Conclusions are summarized in the last part and additional information including the computational details are provided in Appendix.

\section{Structure and charges}\label{sec:LHS_charge_distribution}


The AGBNs studied in this work intercalate armchair nanoribbons of graphene 
with armchair nanoribbons of hexagonal boron nitride, as schematised in Figure~\ref{fig:ag5bn9_structure}. 
They have two characteristic lengths: the number of rows in the graphene ribbon ($N$) and the number of rows in the hBN ribbon ($L$).

Figure~\ref{fig:ag5bn9_structure}.a represents the case where both $N$ and $L$ are odd.
The system is thus characterised by a mirror plane perpendicular to $x$ highlighted with a red solid line.
On the other hand, Figure~\ref{fig:ag5bn9_structure}.b represents the case where $N$ and $L$ are both even.
It has a glide plane ($\bar{b}$/2) perpendicular to $x$ marked with a dashed red line. 
It has to be noted that, for all interfaces to be armchair inside a rectangular unitary cell, $N$ and $L$ must have the same parity, so these two examples fulfill all the possibilities.


In defining the simulation cell (black solid rectangles in Figure~\ref{fig:ag5bn9_structure}), a problem is encountered in accommodating the lattice mismatch~\cite{Nguyen_2019,Lu_2014}.
Indeed, the isolated nanoribbons 
have lattice parameters of 2.45~\AA{ }and 2.50~\AA{ }in Gr and hBN respectively, corresponding to a C-C bond length of 1.42~\AA{ }and a B-N bond length of 1.45~\AA.
Because of this mismatch, in real heterostructures, strain is accumulated along the interface and released locally by the formation of interface defects~\cite{Lu_2014,Akman_2023,Akhtarianfar_2019,Zhang_2022}.
Here, we actually consider perfect interfaces by assembling the two isolated nanoribbons and accommodating the lattice mismatch with a relaxation run.
The main bonds affected by the lattice relaxation are highlighted in Figure~\ref{fig:ag5bn9_structure}.b. 
Internal C-C bonds parallel to the interface are slightly elongated to 1.43~\AA{ } ($+$0.7$\%$) while B-N bonds are contracted to 1.44~\AA{ }($-$0.7$\%$). 
An exception to this behaviour is observed at the interface where C-C and B-N bonds remain  respectively 1.42~\AA{ } and 1.45~\AA{ }long. 
Mixed bonds N-C and B-C are 1.41~\AA{ }and 1.51~\AA{} respectively, in agreement with other studies~\cite{Pruneda_2010}.
Similar deformations are unavoidable whenever defect-free heterostructures are under study and, in this case, they are not expected to impact heavily the electronic properties owing to their small magnitude.
For sake of comparison, similar global longitudinal stretches on isolated nanoribbons~\cite{SerranoRichaud2024} with a width of 5 rows lead to gapwidth reduction of the order of 100 meV in Gr ribbons and an increase of about 40 meV in BN ones with no appreciable change of the charge distribution.

We will release these artificial constraints in a future work devoted to the optical properties of heterostructures in presence of interface defects.



The presence of an interface often comes with a charge redistribution that may have important consequences on the ground- and excited-state properties of a compound.
Figure~\ref{fig:Voronoi_analysis} shows the Voronoi deformation density analysis~\cite{Guerra_2004} reporting the integrated charge on each atom of the unitary cell of a representative heterostructure, but these results can be generalised to structures with different $N$ and $L$ on the condition that both widths are large enough (see section II).
The integrated charges on B, C and N atoms are 
indicated in terms of the excess ($+$) or missing ($-$) charge with respect to the neutral atom.

\begin{figure*}
    \centering
    \includegraphics[width=1.0\linewidth]{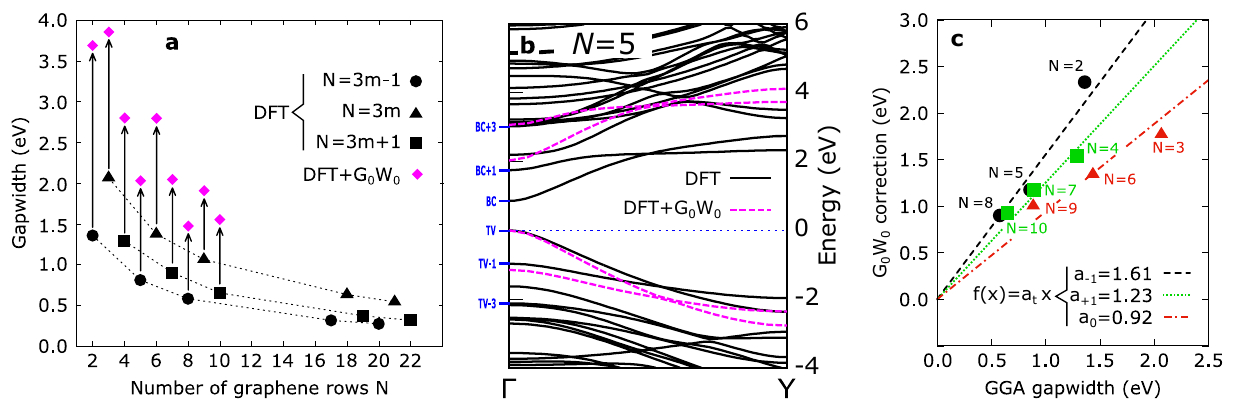}
    \caption{{\bf a}: Evolution of the DFT gap of AGBN as a function of graphene size $N$ (black symbols) and G$_0$W$_0$ corrections (magenta diamonds). Dotted lines are guide for the eyes distinguishing the three families (circles for $N=3m-1$, triangles for $N=3m$ and squares for $N=3m+1$). The size of BN ribbons is either $L$=8 for $N$ even or $L$=9 for $N$ odd. {\bf b}: DFT (solid black) and G$_0$W$_0$ (dashed magenta) band structure close to the gap in the reference heterostructure. Notable states at $\Gamma$ are highlighted with blue tics. {\bf c}: G$_0$W$_0$ corrections to the gap versus DFT gapwidth. Dashed lines highlight a linear fit for each family. Angular coefficients $a_t$ are reported where $t\in\{0,+1,-1\}$ labels the family.}
    \label{fig:dft-gw_gaps}
\end{figure*}

\begin{table}[b]
    \centering
    \begin{tabular}{c @{\hspace{2mm}}|@{\hspace{2mm}} l |@{\hspace{2mm}} c @{\hspace{4mm}} c @{\hspace{2mm}} c }
    \hline \hline 
    & System & DFT & GW & QP corr. \\
    \hline
    \multirow{3}{*}{$m=1$} & AG2BN8  & 1.36 & 3.69 & 2.34 \\
                          & AG3BN9  & 2.06 & 3.86 & 1.79 \\
                          & AG4BN8  & 1.29 & 2.80 & 1.51 \\
    \hline
    \multirow{3}{*}{$m=2$} & AG5BN9  & 0.86 & 2.04 & 1.18 \\
                          & AG6BN8  & 1.44 & 2.80 & 1.36 \\
                          & AG7BN9 & 0.90 & 2.05 & 1.15 \\
    \hline
    \multirow{3}{*}{$m=3$} & AG8BN8  & 0.58 & 1.48 & 0.90 \\
                          & AG9BN9  & 0.89  & 1.91 & 1.03 \\
                          & AG10BN8 & 0.65 & 1.57 & 0.92 \\
    \hline \hline 
    \end{tabular}
    \caption{DFT and DFT+G$_0$W$_0$ band gap at $\Gamma$ in the smallest nine heterostructures grouped by increasing $m$. Values of the AG8BN8 are in reasonably good agreement with those reported in Ref.~\onlinecite{Bernardi_2012b}.}
    \label{tab:gw-dft}
\end{table}

According to our analysis, in the inner rows of the BN part ($i\neq 1$, $L$), each B loses approximately 0.4 electrons equally distributed among the three surrounding N atoms. 
Correspondingly, each N gains about 0.4 electrons from the three surrounding B atoms, as in pristine hBN~\cite{Susana_2024}.
As expected, in the inner rows of the C nanoribbons ($j\neq 1,$ $N$), all C atoms conserve their nominal charge.
At the interface a charge reorganization is indeed observed. 
Each N atom still receives its portion of electrons from the two B atoms that surround it (2/3 of 0.4 $\sim$ 0.27 electrons), and correspondingly the B atoms share their $\sim 0.27$ electrons with the two closest N atoms. 
The interfacial C atoms feel the extra electrons in N and the missing electrons in B. 
Consequently, the edge of the C ribbon polarizes and we observe a charge transfer within the C part of about 0.16 electrons from the N-bound carbons to the B-bound carbons.
In summary, we find no appreciable net charge transfer between hBN and Gr nanoribbons.

Even though this is a known fact in zig-zag graphene/boron nitride interfaces, the issue is not sorted in armchair ones (cfr. Refs.~\mbox{\cite{Seol_2011}} and~\mbox{\cite{Bernardi_2012b}} where opposite claims are made). Here, in agreement with Bernardi et al.~\mbox{\cite{Bernardi_2012b}}, we confirm the absence of charge redistribution across the edge also in armchair interfaces.
In addition, our Voronoi analysis quantifies the charge transfer at the edge of the C ribbons which, to the best of our knowledge, has not been reported yet and which may have an impact on heterostrcuture's chemical activity.

A similar analysis conducted on an $N$-even heterostructure, namely the $N=6$ of Figure~\ref{fig:ag5bn9_structure}b, leads to the same conclusions despite the different symmetry.
In Appendix E, we also report the electron localization function (ELF) plots of the two reference systems of Figure~\ref{fig:ag5bn9_structure}.
They show that electrons basically localised on the C-C bonds inside the Gr region and on the N atoms inside the BN ribbons as a consequence of the different polarity of the C-C and B-N bonds.

\section{Electronic structure of AGBNs}
\label{sec:LHS_elecronic_structure}

\subsection{Gap family classification}

As pointed out in similar works~\cite{Bernardi_2012b,Seol_2011,Guan_2021,Nascimiento_2019}, the size of the C region is crucial to discuss the electronic properties of CBN structures.
This makes sense because one can look at the heterostructure as composed of armchair graphene nanoribbons embedded in BN.

Indeed, the fact that no net charge transfer occurs between the C and the BN parts, partially justifies this pragmatic approach.

So let us start by reporting the bandgap of AGBNs as a function of the graphene width $N$.
We found, in agreement with the literature~\cite{Li_2013,Guan_2021,Nascimiento_2019}, that for C ribbons large enough, the band gap is direct at $\Gamma$ in almost all cases.
However, we observed also some exceptions. \
More precisely we found that those system with  $N+L=3m-1$ display a tendency to develop a direct gap at $X$.
In addition, we found few sparse structures with an indirect gap in the $\Gamma-X$ path, but they seem to occur randomly.
For sake of completeness, the gap in the full list of structures we considered is reported in Table~\ref{tab:gap_AGBN_DFT} in Appendix B.
In any case, the band dispersion is very week in the $\Gamma-X$ direction and negligible in most of the structures, so in the following we will focus on the $\Gamma$ point in all structures.
In Figure~\ref{fig:dft-gw_gaps}a we report the DFT bandgap at $\Gamma$ with black symbols for $N$ ranging from 2 to 22 and $L$ either equal to 8 --- for even $N$ --- or to 9 --- for odd $N$.
As expected~\cite{Bernardi_2012b,Seol_2011,Guan_2021,Nascimiento_2019}, we observe that the gapwidth as a function of $N$ can be classified into three families depending on whether $N=3m$ (family $0$), $N=3m+1$ (family $+1$) or $N=3m-1$ (family $-1$) where $m\in\mathbb{N}^*$.
This characteristic behaviour directly comes from isolated nanoribbons and is a signature of the 1D confinement of the electrons in the armchair direction.
However, in AGBNs, the hierarchy of the three families differs from that of isolated Gr and hBN nanoribbons~\cite{Saraswat_2021,SerranoRichaud2024}. 
In fact, in heterostructures sharing the same value of $m$, the $0$-family hosts larger gaps than the $+1$ family, whereas in isolated Gr and hBN armchair ribbons their order is reversed.
Moreover, if the $-1$ family is globally the lowest-laying gap family both in isolated and heterostructured nanoribbons, in the latter group and only for $m=1$ (i.e. very thin graphene ribbons) $E_g(N=2)>E_g(N=4)$.

In order to have a more reliable quantification of the gapwidth, we computed quasiparticle corrections in the G$_0$W$_0$ approximation. 
In Fig.~\ref{fig:dft-gw_gaps}b we report the DFT band structure of our reference system ($N=5$) together with the GW-corrected dispersion of the near-gap states (magenta dashed lines).   
Unless one has specific needs (e.g. accurate evaluations of the effective mass), we can assert that G$_0$W$_0$ acts pretty much as a rigid shift of the empty states, suggesting the use of a scissor operator as a reasonable approximation. 
This is not surprising in electronic states with a low first quantum number, though the question is how such a shift changes with $N$. 
In Fig.~\ref{fig:dft-gw_gaps}a and Table~\ref{tab:gw-dft}, we compile the GW corrected gap for $N$ ranging from 2 to 10. 
From a quick look at the data, we see that the $N$-dependence of the quasiparticle corrections does not bring any qualitative change to the DFT trend, i.e. the gap hierarchy is the same, however a more careful inspection reveals that corrections themselves obey a three-family classification. 
To highlight this fact, we report the GW correction vs DFT gap in Fig.~\ref{fig:dft-gw_gaps}c and we fit them with the linear regression law
\begin{equation}
    E^{\text{GW}}(N) - E^{\text{DFT}}(N) = a_t \, E^{\text{DFT}}(N)  \,,
\label{eq:linear_so}
\end{equation}
where $E^{\text{GW/DFT}}(N)$ is the GW or DFT gapwidth of the AGBN with a graphene width of $N$ and $t \in \{ 0, +1, -1\}$ labeling the family.
This analysis, besides providing a tool to estimate the GW correction in larger systems where explicit calculations may be unfeasible, provides some insight about the family dependence of electron-electron scattering processes.
In fact, the sequence $a_{-1} > a_{+1} > a_{0}$ can be put in inverse relation with a polarization hierarchy.
For example, the fact that the GW correction to the same DFT gap is smaller in the 0 family than in the other two indicates that electronic screening is more efficient in this family.

At this point, one may wonder why the Gr width dominates the heterostructure gapwidth, or conversely why the gapwidth seems to be insensitive to the hBN width $L$. 
In addition, one can ask whether the width $N$ affects only the gap or other relevant electronic states.
To answer these questions, in the following we carry out a thorough DFT analysis of the states close to the gap.

\subsection{Graphene width and states around the gap}

\begin{figure}[b]
    \centering
    \includegraphics[width=0.50\textwidth]{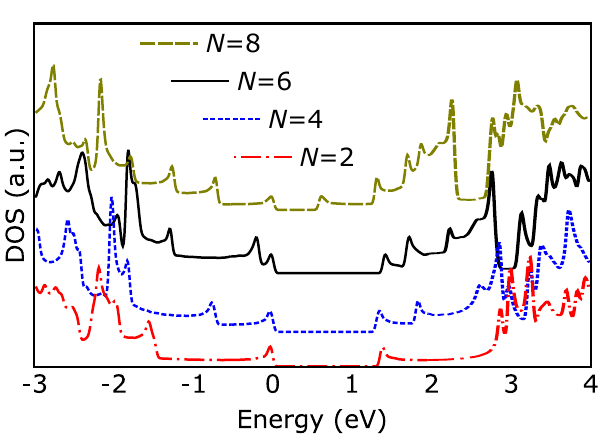}
    \caption{Density of states of AGBN with different $N$ and $L=8$. All curves are aligned with TV at 0 eV.}
    \label{fig:dos_Gr_size}
\end{figure}

Let us 
keep analysing the heterostructure as an array of Gr nanoribbons embedded in hBN.
Following this idea, the role of Gr ribbons can be discussed with regard to two aspects. 
The first is the impact of the Gr width on the electronic properties of the whole heterostructure (current section), the second is the local effect at the interface that modifies the electronic properties of the corresponding isolated ribbon (section II.C). 

To begin with the first point, the role of $N$, let us stress
that if states at $\Gamma$ are non-degenerate at any graphene size $N$, 
the parity of $N$ plays a role at $Y$ where states of all even-$N$ ribbons are doubly degenerate because of the glide symmetry of the system (cfr. Figure~\ref{fig:all_bands} in Appendix C). 
Besides this, we have  plotted in Figure~\ref{fig:dos_Gr_size} the density of states (DOS) for 
$N$ varying from 2 to 8 at a fixed value of $L=8$.
At first glance one can already observe different features between the curves besides the gapwidth.
In particular, the change on the positions of van Hove singularities close to the gap with respect to $N$.
Notice that the erratic trend of the gap is due to the fact that, except for the $N=2$ and $N=8$ curves, the systems belong to different families.

\begin{figure*}
    \centering
    \includegraphics[width=\linewidth]{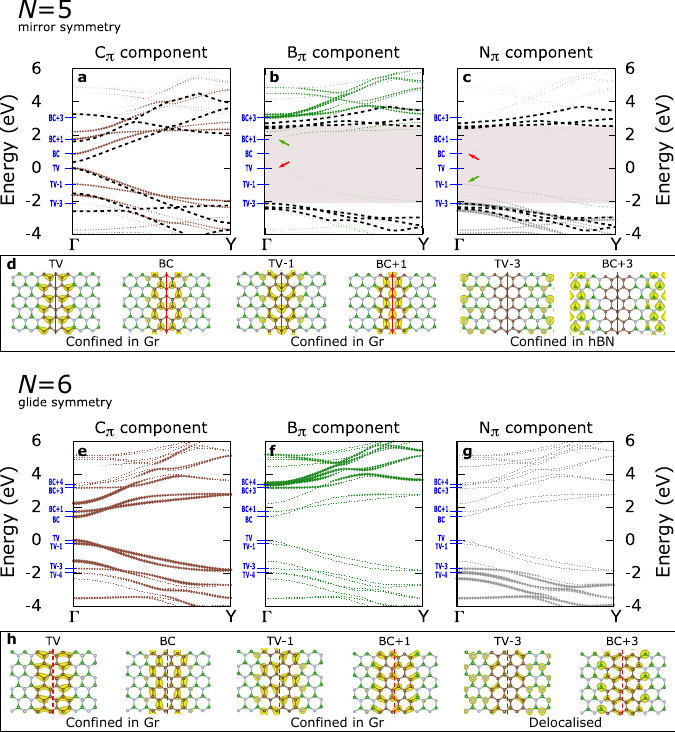}
    \caption{\textbf{a}: DFT band structure of the $N=5$ heterostructure projected onto the $\pi$ orbital components centered on C atoms (C$_\pi$ states). The circle size is proportional to the projected weight. Dashed lines indicate the band structure of the isolated Gr nanoribbon with $N$=5. \textbf{b,c}: The same as \textbf{a} for B$_\pi$ and N$_\pi$ states. Dashed lines are from the isolated BN ribbon with $L$=5. The shaded area corresponds to the DFT gapwidth of the isolated hBN monolayer. Green and red arrows highlight the expected and unexpected hybridization with C states at the interface.
    \textbf{d}: Electronic probability density |$\Psi$($\vc{r}$)|$^2$ at $\Gamma$ of the states TV and BC (top panels), TV$-$1 and BC+1 (middle panels), TV$-$3 and BC+3 (bottom panels) in the $N=5$ AGBN with its mirror symmetry plane marked as a red solid line. \textbf{e}, \textbf{f}, \textbf{g}, \textbf{h}: The same as panels a, b, c and d but in the $N=6$ heterostructure. In d, the glide symmetry plane is marked with a dashed red line.}
    \label{fig:n5m9_projbands}
\end{figure*} 

To have a clearer picture we report in panels a, b and c of Figure~\ref{fig:n5m9_projbands}, the DFT band structure of our reference heterostructure projected onto $\pi$ states --- i.e. combinations of $p_z$ orbitals --- centered on the three different atomic species: from left to right C, B and N components. 
For a collection of band structures including other values of $N$, we refer the reader to Appendix C.
Due to the relative electronegativity of the three atomic species, their band structure components settle in separate energy ranges. 
The bands forming the gap and the closest to it are almost pure C$_{\pi}$ states. 
In particular, they constitute the dominant component of the last two occupied states, i.e. the top valence (TV) and the state just below it (TV$-1$) and the first two empty states, i.e. the bottom conduction (BC) and the one just above it (BC$+1$).
This is consistent with the observation that the gap-family classification is driven by the width of the graphene part.
States far above the gap (about 2~eV above the bottom conduction) are composed mainly of B$_{\pi}$, in particular the first mainly-B state is the fourth empty state BC+3.
On the other side of the gap, deep in the valence (between $-$2~eV and $-$3~eV), the main contributions come from N$_{\pi}$ and C$_{\sigma}$ states. 
In particular, the last state with a dominant N contribution (mainly-N state) is the fourth-last occupied state (labelled TV$-$3). 
We define the energy difference between the mainly-B and mainly-N states at $\Gamma$, with a small abuse of language, the \emph{embedded BN gap} $E^{BN}_\text{emb}$ which in DFT amounts to 5.15~eV.
This distribution is expected, since in the limit of infinite nanoribbons the graphene part should be a semi-metal~\cite{CastroNeto_2009} and hBN a wide gap semiconductor with top valence states centered on N and bottom conduction states on B~\cite{Galvani_2016}. 
For comparison, in the B and N panels (panels b and c), the shaded area represents the DFT gapwidth of the isolated hBN monolayer.

Actually, looking more into details one observes that the mainly-C states forming the gap present a small hybridization with both B$_{\pi}$ and N$_{\pi}$ states highlighted by arrows in panels b and c. 
In agreement with the distribution discussed before, one may expect that the hybridization in conduction is mainly with B$_{\pi}$ states and in valence manly with N$_{\pi}$ states. 
This is indeed the case in the TV$-$1 and BC+1 states, which we highlighted with green arrows. 
However, unexpectedly, the hybridization is reversed in the states composing the gap at $\Gamma$ (red arrows) where the TV presents a weak hybridization with B$_{\pi}$ and the BC with N$_{\pi}$.



To visualise the shape of these states, in Figures~\ref{fig:n5m9_projbands}.d we report the electronic density probability $\left| \Psi_{\vc{k},n}(\vc{r}) \right|^2$ of states with $\vc{k}=\Gamma$ and band index $n$ belonging to bands TV$-$1, TV, BC, BC+1, TV$-$3 and BC+3. 
The maximum of the electronic density of states TV$-$1, TV, BC and BC+1 is localised in the graphene part. 
The confinement in one direction produces the same wave function modulation we reported in armchair Gr nanoribbons~\cite{SerranoRichaud2024}.
However, owing to the presence of BN at the interface, the probability density does not end abruptly at the interface and slightly spreads inside the BN part. 
In particular, we observe a weak protrusion on the N atoms in the TV$-$1 state and on B atoms in the BC+1, and conversely in the TV and BC states. 
The TV$-$2 and BC+2 states present similar characteristics (not shown).
Far from the gap, the electronic density of the TV$-$3 state is centered on the N atoms of the BN ribbon and that of the BC+3 on the B atoms.
Thus in the $N=5$ structure we observe a net passage from states confined in the Gr part and states that are confined in the BN part, if one neglects the weak protrusions at the edges.
All these observations are very consistent with the band decomposition reported in Figures~\ref{fig:n5m9_projbands}a,b, and c.
All the electronic density probabilities satisfy the mirror symmetry of the reference system.

Many of the characteristics discussed above are common to all other structures, in particular the confinement of the near-gap states inside the graphene ribbon even though their number is higher in structures with higher $N$ due to the related piling-up of the bands.
Conversely, other characteristics do not occur systematically in all heterostructures. 
In particular, mixed states in the near-gap energy range can be observed in some systems.
They have a sizeable probability both in the Gr and the hBN parts and are characterised by a high level of hybrdization between the two materials. 
An example is given if Figure~\ref{fig:n5m9_projbands}e, f, g and h where we report the projected band structure and the partial density of notable states in the $N=6$ heterostructure.
The TV$-3$ and the BC+3 states present a pretty flat dispersion and a high level of C-BN hybridization. 
The corresponding charge probability is distributed in both constituent nanoribbons.

\subsection{The impact of the BN interface on Gr ribbons' properties}\label{sec:LHS_Gr_role}

\begin{figure}
    \centering
    \includegraphics[width=\linewidth]{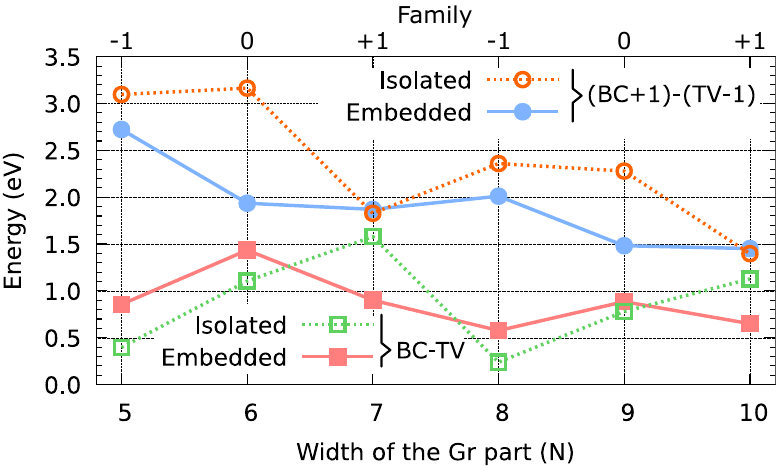}
    \caption{BC-TV and (BC+1)-(TV$-$1) energy gaps at $\Gamma$ (squares and circles) in AGBNs (solid lines) and isolated armchair graphene nanoribbons (dashed lines).}
    \label{fig:levels_lhs-nr}
\end{figure}

To understand how the Gr/BN interface modifies the properties of AGBNs with respect to that of isolated Gr nanoribbons, we shall compare two systems with the same $N$. 
In Table~\ref{tab:embedded_vs_isolated} and Figure~\ref{fig:levels_lhs-nr}, we report the DFT energy of the states TV$-$1, BC and BC+1 in heterostructures and isolated nanoribbons~\cite{SerranoRichaud2024} (with H-passivated edges to saturate dangling bonds) with $N$ ranging from 5 to 10.
The TV states have been aligned to 0 in all systems.
We observe that the classification into three families still holds for the very energies of the near-gap states (not only for the gapwidth), but also, which is more surprising, for the variations with respect to the isolated nanoribbons. 
The latter aspect indicates a family dependence of the response to modifications of the interface.

If we first focus on the gap variations, in family $-1$ the gap is opened in the heterostructure and the same trend is observed in the $0$ family, but with a smaller opening.
At variance, in family $+1$, the BN interface closes the BC-TV gap.
This has the effect of inverting the gap hierarchy of families $0$ and $+1$ in the heterostructure at given $m$ with respect to isolated graphene nanoribbons as it appears clear in Fig.~\ref{fig:levels_lhs-nr}.

To rationalize this gapwidth hierarchy inversion, we rely on the tight-binding (TB) ladder model that has been developed in the past to describe the gap of isolated armchair Gr nanoribbons~\cite{Wakabayashi_1999,SerranoRichaud2024,Son_2006,Seol_2011}.
In this model, edge effects (passivation, stretch, atomic relaxation) can be and have been taken into account through a perturbation Hamiltonian solved perturbatively~\cite{SerranoRichaud2024,Son_2006,Seol_2011} --- resulting in analytic expressions --- or exactly~\cite{SerranoRichaud2024}  via numerical diagonalization.
Following Ref.~\onlinecite{SerranoRichaud2024}, the presence of an interface under negligible edge strain can be included through a change of on-site energies of the edge atoms ($\delta \epsilon_C^{(\text{B})}$ and $\delta\epsilon_C^{(\text{N})}$) to account for their different electrostatic environment. 
A sketch of it is given in Figure~\ref{fig:ladder_lhs}.

\begin{table}[b]
    \centering
    \begin{tabular}{c|@{\hspace{2mm}} c@{\hspace{3mm}} c@{\hspace{3mm}} c@{\hspace{2mm}}|@{\hspace{2mm}}c @{\hspace{3mm}}c @{\hspace{3mm}}c}
    \hline \hline 
    & \multicolumn{3}{ c |@{\hspace{2mm}}}{Embedded }&\multicolumn{3}{c}{Isolated}\\
    $N$ & TV$-$1 & BC & BC+1 & TV$-$1 & BC & BC+1 \\
    \hline
 5  &  $-$0.97  & 0.86 & 1.75 & $-$1.49  & 0.40 &  1.61\\
 6  &  $-$0.19  & 1.44 & 1.75 & $-$1.04 &  1.10 &  2.12\\
 7  &  $-$0.54  & 0.90 & 1.33 & $-$0.18 &  1.58 &  1.65\\
 8  &  $-$0.72  & 0.58 & 1.29 & $-$1.12 &  0.24 &  1.24\\
 9  &  $-$0.28  & 0.89 & 1.20 & $-$0.73 &  0.78  & 1.55\\
 10 &  $-$0.43  & 0.65 & 1.02 & $-$0.16 &  1.13  & 1.24\\
    \hline \hline 
    \end{tabular}
    \caption{Energy with respect to the TV (eV) of the first states close to the gap in the heterostructure (Embedded) and in isolated graphene nanoribbons (Isolated).}
    \label{tab:embedded_vs_isolated}
\end{table}

The perturbative approach actually leads to null gap corrections as a consequence of the transversal mirror symmetry of the isolated nanoribbons' Hamiltonian.
Such a symmetry is actually broken in heterostructures by the alternate contacts with B and N atoms at the edges.
A perturbative approach from isolated nanoribbons is therefore unfit to describe heterostructures unless the symmetry break is included from the very beginning at the unperturbed Hamiltonian level.
Including this asymmetry from the first place is most likely what Seol and Guo managed doing in Ref.~\onlinecite{Seol_2011}, predicting different trends in the $0$ and $+1$ families, but we regretfully have not been able to derive their Equation 1. 
On the other hand, the exact numerical diagonalization of the ladder model can handle such a symmetry breaking in an exact way and in all families~\cite{SerranoRichaud2024}.
The approach (summarized in Figure 7 of the Ref.~\onlinecite{SerranoRichaud2024}) predicts a quadratic dependence of the gapwidth on $\delta \epsilon_{C}^{(\text{B,N})}$ in all families, though with different signs.
In families $0$ and $-1$ the quadratic coefficient is positive, indicating that the gap should open because of the interface, while in family $+1$ the coefficient is negative, indicating that the gap should close.
This opposite quadratic dependence is responsible of the hierarchy inversion when graphene nanoribbons are interfaced with BN which we observe in our DFT calculations

As it will be pertinent for the discussion on absorption spectra, we now extend this analysis to the transition from TV$-$1 to BC+1.Also in this case a classification into three families is relevant not only to discuss the energy of each state, but also the deviation from the isolated nanoribbon case.
In particular, we find that the hierarchy of the latter differ from that of the TV-to-BC transitions as it goes in the opposite direction in families $-1$ and $0$ and is almost negligible in family $+1$.

In summary, this analysis leads to two results. 
First, we report the evolution as a function of $N$ of the first electronic transition beyond the gap and show that it has a different trend with respect to the latter.
Second, by comparing with isolated nanoribbons, we have shown that the (TV$-$1)-to-(BC+1) and TV-to-BC transitions react  to modifications of the interface in different and sometimes opposite ways.
This insight is particularly relevant whenever interface effects are at study.

\subsection{The role of hBN in the heterostructure}\label{sec:LHS_BN_role}


At this point is worth asking if the hBN region impacts in any way the heterostructure electronic properties.
To this aim, we now look at the AGBNs as arrays of hBN nanoribbons embedded in Gr and we discuss the changes in the B and N states with respect to hBN isolated nanoribbons of the same width. 
Once again we  focus on the effects at the interface and the impact of the width.

\begin{figure}
    \centering
    \includegraphics[width=1.0\linewidth]{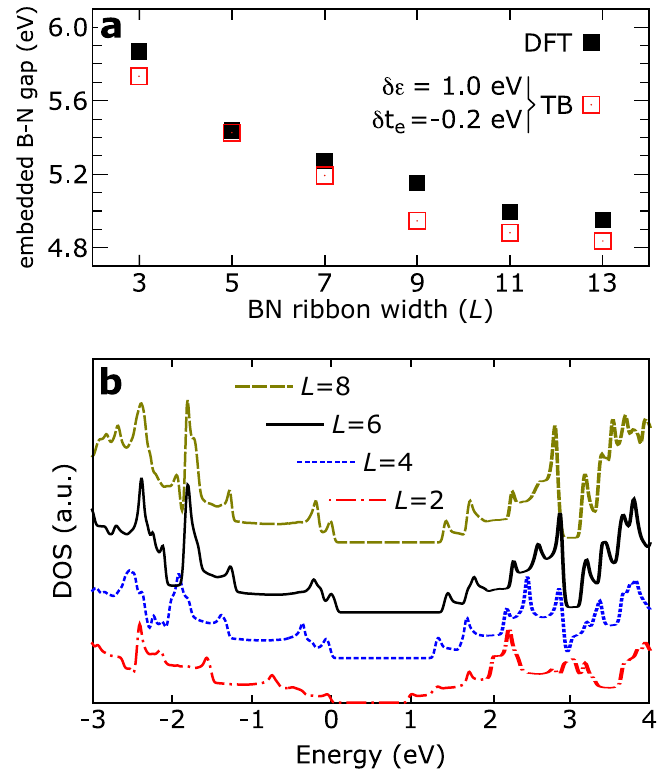}
    \caption{\textbf{a}: Embedded B-N gapwith of AGBNs with $L$ ranging from 3 to 13 within DFT in black solid squares and within numerically solved TB ladder model in red squares. \textbf{b}: Density of states of AGBNs with $L$ ranging from 2 to 8 at $N$=6.}
    \label{fig:dos_BN_size}
\end{figure}

To discuss the first point, let us go back to Figures~\ref{fig:n5m9_projbands}.b and~\ref{fig:n5m9_projbands}.c, where we report in dashed black lines the DFT band structure of the corresponding isolated BN nanoribbon with $L=9$.
For the comparison to be meaningful, we aligned the top valence of the latter to the TV$-$3 state, i.e. the last mainly-N occupied state of the $L=9$ heterostructure, which is at $-2.11$~eV.
This comparison shows clearly how embedding this hBN ribbon in Gr opens its gap by about 600 meV.
This trend is in agreement with the exact diagonalization of the ladder model discussed in the previous section as family $-1$ opens the gap.
Notice that, contrary to the Gr case, in BN  the perturbative approach can be  applied safely  because the unperturbed Hamiltonian of the isolated nanoribbon has the same symmetry as the embedded one.
In Figure~\ref{fig:dos_BN_size}.a, we report  the embedded gap $E_\text{emb}^{BN}$ of AGBN with $N=5$ and $L$ ranging from 3 to 13 from DFT (solid black) and the exact diagonalization TB (empty red).
The TB parameters we use here are the same as in the isolated (H-passivated) nanoribbons~\cite{SerranoRichaud2024}, except for the perturbation. 
In fact, to account for the bounding with C instead of H, here the on-site perturbation is set to $\delta\epsilon_1= - \delta\epsilon_2=\delta\epsilon=1.0$~eV instead of $-0.145$~eV and the edge bond stretching $\delta t_e = -0.2$~eV instead of  $-0.19$~eV. 
Both parameters have a higher absolute value in the embedded system than in the isolated one, displaying the strong impact of C at the interface. 
The agreement between DFT and the ladder model demonstrates that the opening of $E_\text{emb}^{BN}$ in the heterostructure does come from an edge perturbation due to the C atoms.

Let us now pass to the second point. 
To illustrate the impact of the BN width $L$ on the electronic properties of AGBN, we plot the DFT DOS of $N=6$ heterostructures for different values of $L$. 
In Figure~\ref{fig:dos_BN_size}.b, $L$ varies from 2 to 8.
The AGBNs with the largest BN ribbons ($L$=6 and $L$=8) present the same DOS 2~eV below and above the Fermi level. 
They have a DFT gapwidth of 1.44~eV and the van Hove singularities around the gap display the sharp peaks characteristic of a quasi-1D system.
We conclude that in these systems interactions between Gr nanoribbons are negligible, i.e. 6 BN rows are enough to isolate the Gr ribbons in this energy range. 
This also implies, in agreement with Ref.~\onlinecite{Guan_2021}, that larger BN ribbons have no impact on states close to the gap. 
Because of this, in this study, the width of the BN ribbons is set either to $L$=8 (for $N$ even) or $L$=9 (for $N$ odd).

\bigskip

In conclusion to this part, we have determined the specific role of each material on the electronic properties of AGBNs.
The tuneability of the heterostructure electronic properties around the Fermi energy is entirely related to the Gr width $N$, while the differences between isolated and embedded armchair nanoribbons come from local perturbations due to the interface.
This latter point can be understood by perturbing a tight-binding ladder model and suggests that any additional perturbation at the interface, such as defects, 
will affect near-gap states in different and sometimes opposite fashion. 
In addition, we scrutinized quasiparticle corrections and put forward that the efficiency of the electronic screening grows in a family sequence $-1$, +1, 0.
As a by-product of this analysis we propose a linear relation between DFT gapwidth and QP correction which can be of interest for further studies on larger and more complex AGBN structures.


\section{Absorption spectra}\label{sec:LHS_absorption}

So far we have investigated the electronic structure of AGBN, relying on DFT simulations and on a model to understand it in particular with respect to isolated Gr or hBN nanoribbons.
In this section, we focus on the spectral properties of AGBNs in the independent particle approximation (IP) looking at them from the same perspective.
To start with, we first derive how selection rules change from those of isolated AGNRs and then we discuss the characteristics of absorption spectra in each family.

\subsection{Selection rules}\label{sec:LHS_selection_rules}

Let us recall that if $N$ and $L$ are large enough (approximately $N \ge 3$ and $L \ge 6$), the gap of AGBNs will mainly be due to $\pi$ states localised on the Gr part. 
Then one can assume that the onset of an AGBN absorption spectrum is dominated by transitions whithin C-centered states. 
However, the interface with BN breaks the transversal mirror symmetry of isolated ribbons. 
In the following, we show how this loss of symmetry can harbour new optically active transitions. 
To this aim, we will calculate the selection rules of this system for transitions taking place at the $\Gamma$ point.
Contrary to similar derivations~\cite{Nematian2012}, in this work we revert to the ladder model presented in our previous work~\cite{SerranoRichaud2024} with the intent of highlighting the different behaviour with respect to isolated nanoribbons.

We adopt here the same notation as in Ref.~\onlinecite{SerranoRichaud2024}, apart from $N_a$ here replaced by $N$.
In the following derivation, the higher symmetry of the unperturbed Hamiltonian will not lead to an artificial cancellation of terms because there is no energy difference involved in the calculation as opposed to the gapwidth corrections derived above.
So we can seek for a  meaningful analytical expression within first order perturbation theory.

\begin{figure}
    \centering
    \includegraphics[width=0.90\linewidth]{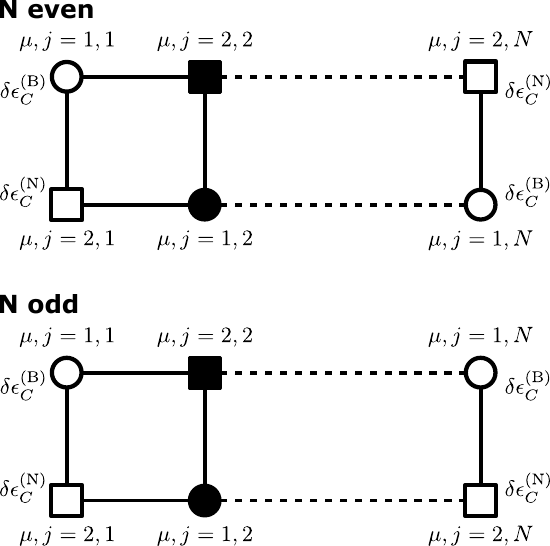}
    \caption{Sketch of the indexing convention and interface perturbation $V$.}
    \label{fig:ladder_lhs}
\end{figure}

The perturbation Hamiltonian $V$ contains on-site energy variations of interface carbons $\delta \epsilon _{\mu j}$ that are due either to a C-B bond ($\delta \epsilon _{C}^{(\text{B})}$) or a C-N bond ($\delta \epsilon _{C}^{(\text{N})}$).
Here $\mu$ labels the sublattice, and $j$ the row of the ladder.
Therefore, if $j$ is odd, $\mu=1$ labels sites of the top stringer and if $j$ is even, it labels sites of the bottom stringer.
The opposite holds for $\mu=2$. 
With this convention, the edge perturbations do not depend on the parity of $N$ and are 
\begin{equation}
    \delta\epsilon_{11} = \delta\epsilon_{1N} = \delta\epsilon_C^{(\text{B})}
    \quad\text{and}\quad
    \delta\epsilon_{21} = \delta\epsilon_{2N} = \delta\epsilon_C^{(\text{N})}\,.
    \label{eq:V}
\end{equation}
For sake of clarity, we sketch this setup in Figure~\ref{fig:ladder_lhs}.

In first order perturbation theory, the eigenstates are  
\begin{align}
    \ket{n\pm}&=\ket{n\pm,0} + \ket{n\pm,1} \nonumber \\
    & = \ket{n\pm,0} +  \sideset{}{'}\sum_{m,q}\frac{{\bra{mq,0}}V\ket{n\pm,0}}{E_{n\pm}^0-E_{mq}^0}\ket{mq,0} \,.\label{eq:wfc_perturbed_general}
\end{align}
In this expression $E_{n\pm}^0$ are the eigenvalues of the unperturbed ladder model, $q$ labels either empty ($+$) or occupied ($-$) states, the primed sum excludes $\ket{n\pm,0}$ from the summation, and
\begin{equation}
\ket{n\pm,0} = \sqrt{\frac{2}{N+1}} \sum_{j=1}^N \sum_{\mu} \sin\left( j \theta_n \right) D_\mu^{n\pm} \ket{\mu,j}
\label{eq:n0}
\end{equation}
 are the unperturbed eigenstates.
We recall that $\theta_n = n\pi(N+1)^{-1}$ comes from the quantization of the momentum.
For the purpose of the current study, it is not necessary to know the exact experssion~\cite{SerranoRichaud2024} of the sublattice coefficients $D_\mu^{n\pm}$.

The semiclassical expression of the optical absorption of a gapped system at a frequency $\omega$ in the IP framework is reported in equation~\cite{Grosso2013}: 
\begin{equation}\label{eq:Abs}
     A(\omega) \propto \sum_{n,m}  \text{Im} \left[ \frac{\left| \mathcal{M}_{nm} \right|^2}{E_{n+}-E_{m-} -\omega -i\eta} \right] \,,
\end{equation}
where $\mathcal{M}_{nm}=\bra{n+}\vc{e} \cdot \textbf{v} \ket{m-}$ is the velocity matrix element with $\vc{e}$  the polarization vector of the electromagnetic field,  $\textbf{v}$ the velocity operator
and $\eta$ an infinitesimal parameter to remove poles from the real axis to ensure analyticity of the expression.
As the quasi 1D confinement quenches the absorption in the direction perpendicular to the ribbon, we only develop the selection rules with light polarized along the ribbon axis, $\vc{e}=\vc{e}_y$. 
Then, the  velocity matrix element at the first order in the perturbation are:
\begin{equation}\label{eq:Mnm_perturbed}
\begin{split}
        \mathcal{M}_{nm} =\mathcal{M}_{nm}^0 &+\bra{n+,1}\text{v}_y\ket{m-,0} +\\
        &+\bra{n+,0}\text{v}_y\ket{m-,1}\, , 
    \end{split}
\end{equation}
where $\mathcal{M}_{nm}^0$ is the known velocity matrix of the unperturbed system~\cite{Chung_2011}. 
Because of the confinement in the armchair direction, $\mathcal{M}_{nm}^0$ allows only transitions between states with the same $n$, namely:
\begin{equation}\label{eq:Mnm_unperturbed}
    \mathcal{M}_{nm}^0 = \bra{n+,0} \text{v}_y \ket{m-,0}  = \mathcal{P}_{nn} \delta_{nm} \,,
    \end{equation}
where $\mathcal{P}_{nn}$ is the oscillator strength of the (allowed) transition.
We remind that $E_{n\pm}^0$, although not necessarily ordered with respect to the gap, are symmetrically distributed around it which eases the correspondence with the band structure notation we have used before.
The equation above states that the only optically active transitions in unperturbed nanoribbons are those from TV to BC, from TV$-$1 to BC+1, etc~\cite{Chung_2011}.

The second and third terms of equation~\eqref{eq:Mnm_perturbed} are first order corrections to the velocity matrix.
The main goal of this section is to check if first order corrections may occasion new optically active transitions, i.e. if there is any pair $n\neq m$ such that $M_{nm} \neq 0$.
The second term of equation~\eqref{eq:Mnm_perturbed} reads
\begin{equation}\label{eq:Mnm_1correction}
\begin{split}
    &\bra{n+,1} \text{v}_y \ket{m-,0} =\\ &=\sideset{}{^\prime}  \sum_{k,q} \frac{\bra{n+,0}V\ket{kq,0}}{E_{n+}^0-E_{kq}^0} \bra{kq,0} \text{v}_y \ket{m-,0} \, .
    \end{split}
\end{equation}
Using \eqref{eq:Mnm_unperturbed}, we end up with only two terms, one containing $\bra{m+,0} \text{v}_y \ket{m-,0}$ and the other $\bra{m-,0} \text{v}_y \ket{m-,0}$.
The latter is actually the expectation value at $\Gamma$ of the group velocity of $\ket{m-,0}$ which is a non-degenerate stationary state of the unperturbed system.
As a consequence, its group velocity is null by definition and we are left with one single term leading to the expression
\begin{equation}\label{eq:Mnm_1correction_vs2}
    \bra{n+,1} \text{v}_y \ket{m-,0} = \frac{\bra{n+,0}V\ket{m+,0}}{E_{n+}^0-E_{m+}^0} \mathcal{P}_{mm}\, .
\end{equation}
The same analysis is valid for the other correction term, which reads
\begin{equation}\label{eq:Mnm_2correction_vs2}
    \bra{n+,0} \text{v}_y \ket{m-,1} = \frac{\bra{n-,0}V\ket{m-,0}}{E_{m-}^0-E_{n-}^0} \mathcal{P}_{nn}\, .
\end{equation}

\begin{widetext}
Introducing these expressions into equation~\eqref{eq:Mnm_perturbed} for $n \neq m$, and remembering that $\mathcal{M}^0_{nm}=0$ because of \eqref{eq:Mnm_unperturbed}, we get:
\begin{equation}\label{eq:Mnm_perturbed_vs2}
        \mathcal{M}_{nm}  = 
        \frac{\bra{n+,0}V\ket{m+,0}}{E_{n+}^0-E_{m+}^0} \mathcal{P}_{mm} +\frac{\bra{n-,0}V\ket{m-,0}}{E_{m-}^0-E_{n-}^0} \mathcal{P}_{nn} \,.
\end{equation}

Whether $\mathcal{M}_{nm}$ is null or not, depends on the values of the perturbation matrix elements $\bra{n\pm,0}V\ket{m\pm,0}$ that can be explicitly obtained using~\eqref{eq:n0} and the definition of $V$, i.e. formulae \eqref{eq:V} and their explications.
This leads to:
\begin{equation}
        \bra{n\pm,0}V\ket{m\pm,0}
         = \frac{2}{N+1} \sin(\theta_{m})\sin(\theta_n) 
        \left[ \delta \epsilon _C^{(\text{B})} D_1^{m\pm}D_1^{n\pm}+ \delta \epsilon _C^{(\text{N})} D_2^{m\pm}D_2^{n\pm}\right]\mathcal{F}(n,m) \,,
        \label{eq:selection_rule}
\end{equation}
where the function $\mathcal{F}(n,m) = [ 1+(-1)^{m+n} ] $ 
sets to zero all terms where $m$ has not the same parity as $n$.
To conclude, please note that none of these conclusions prevent $\mathcal{M}_{nm}$ from being zero for other different reasons. 

\end{widetext}

\begin{figure*}
    \centering
    \includegraphics[height=0.90\textheight]{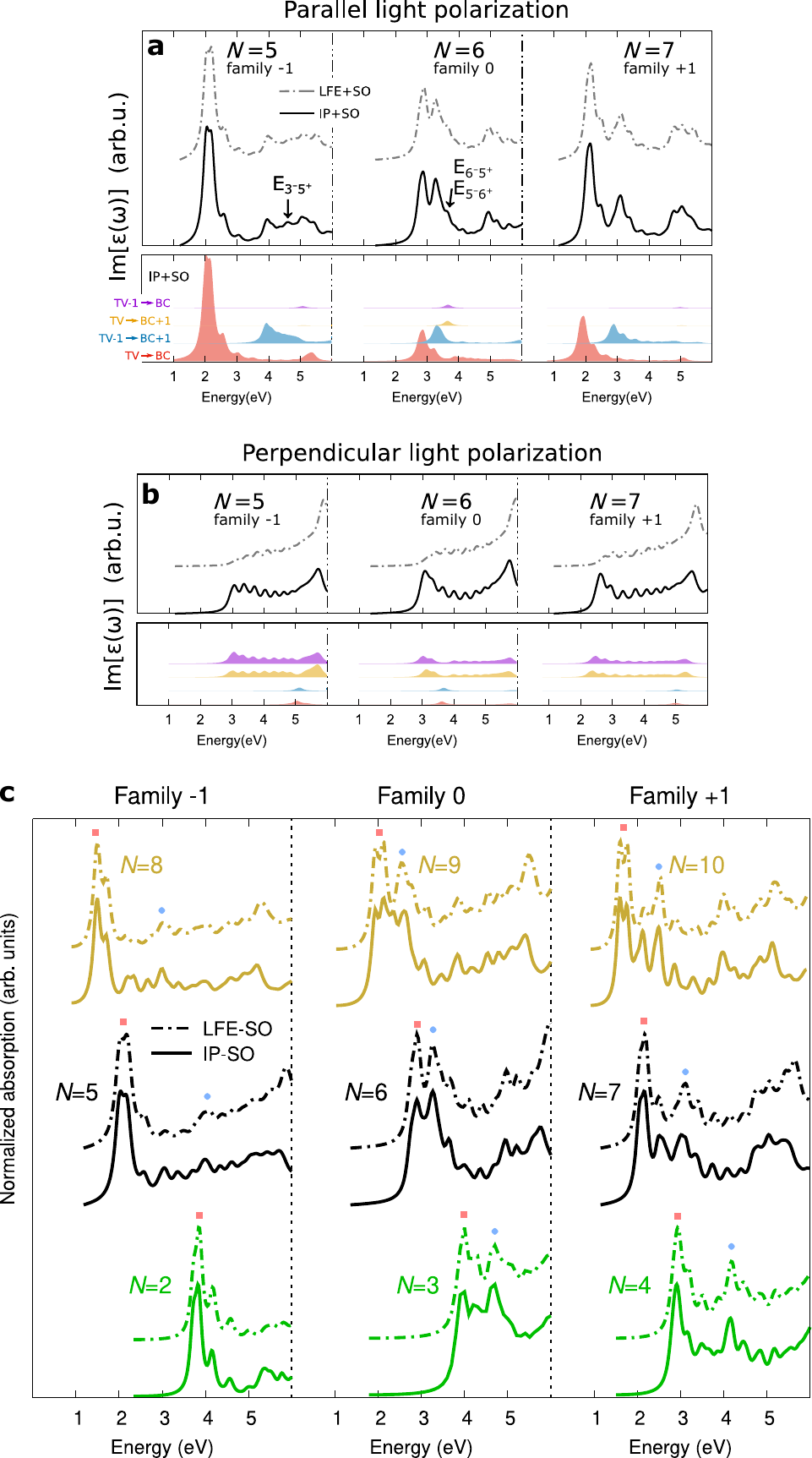}
    \caption{\textbf{a}: Spectra of the $N$=5, 6 and 7 structures from left to right with light polarized parallel to the nanoribbon axis. Top panels: TD-DFT calculations in the RPA including local fields effect (LFE+SO in dotted-dashed grey) and neglecting it (IP+SO in black solid). Arrows indicate notable transitions (see text). Bottom panels: Transition-by-transition decomposition of the IP spectra into the four components TV$-$1$\rightarrow$BC (purple), TV$\rightarrow$BC+1 (yellow), TV$-$1$\rightarrow$BC+1 (blue) and TV$\rightarrow$BC (red).
    \textbf{b}: As \textbf{a}, but for light polarized perpendicular to the ribbon.
    \textbf{c}: Spectra for light polarized linearly at 45$^\circ$ with the ribbons' axis.
    Spectra are grouped vertically by family and horizontally by $m$ (green $m=1$, black $m=2$, yellow $m=3$).
    Primary peaks of LFE+SO spectra are highlighted with a red rectangle, secondary peaks with a blue circle. }
    \label{fig:spectra_rpa}
\end{figure*}


\subsection{Spectral properties and transition analysis}\label{sec:LHS_optical_properties}

Absorption spectra are calculated here from first principles to analyse the effects of the interface from a computational perspective.
 
Spectra are computed in the time-dependent DFT  framework with transition energies corrected with a scissor operator (SO) matching the G$_0$W$_0$ corrections to the
gapwidth.
The polarizability is computed in the random phase approximation (RPA) with and without local field effects (LFE).
We label the former calculations LFE+SO and the latter IP+SO.

Figure~\ref{fig:spectra_rpa} shows representative spectra for each family with a 
field polarization parallel (panel a) or perpendicular (panel b) to the ribbon axis.
We will start discussing the TD-DFT spectra with polarization along the ribbon.

The IP+SO are drawn with a black solid line. 
The difference with grey dashed spectra is the inclusion of local field effects (LFE+SO).
One can easily recognise that the impact of the LFE is negligible and the imaginary part of the dielectric function, $\text{Im}(\epsilon)$ 
is well described at the IP level. 
In the bottom panel, the shaded areas are the partial IP+SO spectra obtained by considering separately transitions between the four bands closer to the gap: TV$-$1, TV, BC and BC+1.
As the selection rules of the unperturbed system~\cite{Chung_2011} 
dictate, the onset of the absorption spectra are composed mainly of transitions between bands symmetrically distributed with respect to the gap. 

The onset and the most intense signature come from the transition from TV to BC at $\Gamma$ (red areas).
The shoulders on the right of this first peak are artificial features due to the coarseness of the k-point grid. 
At higher energy a second less intense peak appears (blue area) which comes from transitions from TV$-$1 to BC+1 principally at $\Gamma$. 
As already pointed out in discussing Figure~\ref{fig:levels_lhs-nr}, the distance between these two main peaks is characteristic of each $N$ family. 
In the $N$=5 (family $-1$), they are separated by about 2~eV. 
In the $N$=6 (family $0$), they overlap in the total spectrum (0.5~eV) and in the $N$=7 (family $+1$), they are separated by about 1~eV. 
As depicted in Figures~\ref{fig:levels_lhs-nr} and~\ref{fig:spectra_rpa}.c, the difference in energy between the two peaks decreases with $N$ inside each family.
However, for the same $m$ they keep the hierarchy detailed above.

Small arrows highlight features of the spectra where $\mathcal{M}_{nm}$ with $n \neq m$ has a sizeable weight. 
Energy labels $E_{n^-m^+}$ report the $n$ and $m$ indexes of the corresponding states $\ket{n-}$ and $\ket{m+}$ of the ladder model. 
In the $N$=5, the label $E_{3^-5^+}$ indicates a transition between odd states, in agreement with the selection rule~\eqref{eq:selection_rule}. 
Here, the ladder model indexing  corresponds to DFT band ordering TV$-$1$\rightarrow$BC+2.
In fact, the link between ladder model index, and DFT band indexing is not always trivial and is given in Appendix D.
In the $N$=6 system, there are signatures of transitions $\ket{5-}\rightarrow\ket{4+}$ and $\ket{4-}\rightarrow\ket{5+}$, i.e. states with different parity. 
In the band structure labeling, these transitions go from from TV to BC+1 and to TV$-$1 to BC (cfr. Figure~\ref{fig:spectra_rpa}.a and see Appendix D for the link between $n\pm$ and band indexing), but they are not occurring in $\Gamma$, so they do not violate the selection rules derived above.



A similar analysis for light polarized perpendicular to the ribbon is presented in Figure~\ref{fig:spectra_rpa}.b.
As the onset of the spectrum is dominated by C$_\pi$ states, the spatial confinement in the light polarization direction produces a decrease in the spectral intensity although at the level of 
IP+SO is not quenched to zero. 
The total quenching of the spectral onset appears at the LFE+SO level which takes into account the inhomogeneity of the heterostructures in this direction. 

We can now discuss absorption spectra in a larger range of graphene widths.
In Figure~\ref{fig:spectra_rpa}.c, we report the LFE+SO and the IP+SO spectra for $N$ ranging from 2 to 10, i.e. for three systems per family under a light polarised linearly at 45$^\circ$ with respect to the the ribbons' axis.
Even though details of the spectra differ from one system to the other, we can observe family-specific characteristics. 
Let us first focus on LFE-SO spectra (dotted dashed lines).
Spectra of the $-$1 family display a primary peak as a rather sharp structure at the onset followed by a quite extended dip that eventually grows forming a relatively weak secondary peak.
The primary peak is highlighted with a red square, indicating that it is mainly due to TV to BC transitions, while the secondary peak, highlighted with a blue circle, is mostly due to TV$-$1 to BC+1 transitions. 
Note that in the $N$=2 system, the latter peak arise at a too large energy to appear in the plot.
The spectral onset in structures belonging to the 0 family is conversely characterised by a single quite broadened feature topped off with a primary and a secondary peak of comparable intensity.
This is because the two main sets of transitions (TV to BC and TV$-$1 to BC+1) are much closer in energy and have comparable oscillator strengths. 
Finally, spectra belonging to the +1 family fall in between the other two families.
In fact these spectra are again characterised by a primary and a secondary peak separated by a dip, as in the $-$1 family, but with smaller differences in energy and intensities.
Besides these features, additional peaks between the primary and the secondary ones may be observed in the IP-SO spectra, but they must be ascribed to perpendicular components that, as we have shown before, are washed out when local field effects are included.
Before concluding we stress how these family-specific spectral features are perfectly consistent with the energy differences scrutinised when discussing Figure~\ref{fig:levels_lhs-nr}.

\section{Conclusion}

To conclude, we have carried out a consistent and comprehensive study of the electronic properties of armchair graphene/boron nitride lateral heterostructures using ab initio methods and models.

We found no net charge transfer at the armchair interface though a local charge redistribution appears at the edge of the graphene ribbons.
We show that these heterostructures (almost always) present a direct gap at $\Gamma$ composed mainly of C-centered $\pi$ states, in agreement with literature.
The use of a tight-binding (TB) ladder model originally introduced for graphene nanoribbons allowed us to gain insight on the way each material and their interface determine the electronic properties of the heterostructure.
Thus, we have concluded that the tunability of the gap relies essentially on the graphene width, while the interface with hBN plays a role in renormalizing energy gaps with respect to the isolated ribbons.

We also have shown that classifying these systems into three families according to the graphene ribbons' width is relevant not only for the gapwidth, as it is known in literature, but also for other properties including near-gap transition energies and their response to modifications of the interface, electronic screening (disclosed by three linear trends in the GW corrections) and spectral peak intensity close to the onset, calculated within the random phase approximation (RPA).
Besides, we have indicated and discussed the trend of all these quantities as a function of the graphene's width.



Finally, the optical selection rules we derived from TB  together with the explicit calculation of absorption, revealed a richer spectrum than that of constituent nanoribbons.
In this respect, the choice of calculating optical properties also at the independent particle (IP) level allowed us to bridge the electronic structure and spectra, permitting a clear identification of onset peaks.

All in all, our study builds a solid foundation for more refined investigations on the absorption properties of graphene/boron nitride lateral heterostructures that may deal with excitonic properties or with the effect of point defects.

\section*{Appendix A: Computational details}

DFT calculations are carried out within the generalized gradient approximation using the Perdew-Burke-Ernzerhof~\cite{Perdew_1996} exchange correlation potential (PBE) as implemented in the Quantum ESPRESSO simulation package~\cite{QE-2009,QE-2017}.
To avoid interactions between consecutive cells, we include 15~\AA{ }of empty space in the $z$ axis.
We have used norm-conserving pseudopotentials~\cite{Hamann_2013} and set the kinetic energy cutoff at 80 Ry. 
The k-point grids employed in this study are 4$\times$8$\times$1 centred on $\Gamma$ to calculate the self-consistent charge density and 4$\times$30$\times$1 for the DOS.
All DOS spectra are convoluted with a Gaussian broadening with a width of 0.02~eV.
Structural relaxation is performed using the BFGS algorithm allowing for optimization of both cell parameters and atomic positions. 
The stopping criterion is that all force components are lower than $5 \times 10^{-5}$~eV/\AA.

Absorption spectra in the random phase approximation with local fields (LFE+SO) and without (IP+SO) are calculated with the Yambo package~\cite{Marini_2009}.
For spectra, we use the same k-point grid as for the DOS, except in the $N=2$, 3 and 4 where a k-sampling of 6$\times$30$\times$1 is necessary.
The energy cutoff of the dielectric matrix is set to 4~Ry and the number of conduction bands included to calculate the dipole matrix elements is twice the number of valence bands.
The DFT energies of the empty states have been shifted with a scissor operator obtained from the quasiparticle gap.
All these parameters ensure convergence of the spectra up to 20~eV.
Finally, all spectra are convoluted with a Gaussian broadening with a width of 0.12~eV. 

Quasiparticle corrections to the last two valence and first two conduction bands are calculated using the Yambo package~\cite{Marini_2009} within the G$_0$W$_0$ approximation.
The self-energy cutoff is 80~Ry.
The number of bands included in the calculation of the self-energy using the band terminator of Bruneval and Gonze~\cite{Bruneval_2008} is eight times the total number of occupied bands.
The screening potential is based on the Godby-Needs plasmon pole approximation~\cite{Rojas_1995} computed at an imaginary frequency of 17.5~eV.
For the calculation of the screening matrix we use the same number of bands and the same k-point grid as for the absorption calculations, while the size of the matrix is increased via a cutoff energy of 5~Ry. These parameters ensure convergence of quasiparticle corrections with an error below 300~meV.

\section*{Appendix B: The role of global width in the heterostructure}

\begin{table}
  \centering
    \begin{tabular}{c @{\hspace{3mm}} c @{\hspace{3mm}}c @{\hspace{4mm}} c @{\hspace{3mm}} l }
    \hline
       $N,L$ &$\Delta_X$ &$\Delta_\Gamma$  &  k-points   & $N$ + $L$    \\ \hline
        \hline
        2,8  & 1.410 & 1.356 &  $\Gamma \to \Gamma$   & 3m+1  \\ 
        3,9  & 2.101 & 2.065 &  $\Gamma \to \Gamma$   & 3m  \\ 
        4,8  & 1.389 & 1.289 &  $\Gamma \to \Gamma$   & 3m  \\
        \hline
        \hline
        5,7  & 0.871 & 0.793 & $\Gamma \to \Gamma$    & 3m  \\
        5,9  & 0.809 & 0.856 & $\Gamma \to \Gamma$ & 3m$-$1   \\
        5,11 & 0.823 & 0.809 & $\Gamma \to \Gamma$ & 3m+1  \\
        5,13 & 0.838 & 0.838 & $\Gamma \to X$    & 3m  \\
        5,15 & 0.839 & 0.841 & $X \to X$ & 3m$-$1  \\
        \hline
        \hline
        6,6  & 1.466 & 1.343 & $\Gamma \to \Gamma$ & 3m  \\ 
        6,8  & 1.374 & 1.435 & $X \to X$ & 3m$-$1  \\  
        6,10 & 1.416 & 1.402 & $\Gamma \to \Gamma$ & 3m+1  \\
        6,12 & 1.412 & 1.414 & $\Gamma \to \Gamma$ & 3m  \\
        \hline
        \hline
        7,9  & 0.901 & 0.901 & $X\to\Gamma$ & 3m+1   \\ 
        7,11 & 0.900 & 0.889 & $\Gamma \to \Gamma$ & 3m  \\
        7,13 & 0.888 & 0.894 & $X \to X$  &3m$-$1  \\
        \hline
        \hline
        8,8  & 0.604 & 0.577 & $\Gamma \to \Gamma$ & 3m+1  \\ 
        8,10 & 0.603 & 0.597 & $\Gamma \to \Gamma$ & 3m  \\
        8,12 & 0.601 & 0.676 & $X \to X$ & 3m$-$1  \\
        \hline
        \hline
        9,9  & 0.906 & 0.886 & $\Gamma \to \Gamma$ & 3m  \\ 
        9,11 & 1.069 & 1.080 & $X \to X$ & 3m$-$1  \\
        9,13 & 1.084 & 1.080 & $\Gamma \to \Gamma$  & 3m+1 \\
        \hline
        \hline
        10,8  & 0.697 & 0.650 & $\Gamma \to \Gamma$ & 3m  \\ 
        10,10 & 0.659 & 0.677 & $\Gamma \to \Gamma$ & 3m$-$1  \\
        10,12 & 0.662 & 0.660 & $\Gamma \to \Gamma$ & 3m+1 \\
        \hline
        \hline
        17,9  & 0.315 & 0.333 & $X \to X$ & 3m$-$1  \\ 
        17,11 & 0.332 & 0.327 & $\Gamma \to \Gamma$ & 3m+1  \\
        17,13 & 0.333 & 0.334 & $\nicefrac{1}{3} \to \Gamma$ & 3m \\
        \hline
        \hline
        18,8  & 0.605 & 0.632 & 
        $X \to X$ & 3m$-$1  \\ 
        18,10 & 0.629 & 0.623 & $\Gamma \to \Gamma$ & 3m+1 \\
        18,12 & 0.632 & 0.632 & $\Gamma \to \Gamma$  & 3m  \\
        \hline
        \hline
        19,9  & 0.374 & 0.372 & $\nicefrac{1}{4} \to \Gamma$ & 3m+1  \\ 
        19,11 & 0.370 & 0.365 & $\Gamma \to \Gamma$  & 3m  \\
        19,13 & 0.361 & 0.363 & $X \to X$ & 3m$-$1  \\
        \hline
        \hline
        20,8  & 0.286 & 0.273 & $\Gamma \to \Gamma$ & 3m+1  \\ 
        20,10 & 0.285 & 0.283 & $\Gamma \to \Gamma$ & 3m  \\
        20,12 & 0.287 & 0.290 & $X \to X$ & 3m$-$1  \\
        \hline
        \hline
        21,9  & 0.550 & 0.542 & $\Gamma \to \Gamma$ & 3m  \\ 
        21,11 & 0.550 & 0.555 & $X \to X$ & 3m$-$1  \\
        21,13 & 0.559 & 0.558 & $\Gamma \to \Gamma$ & 3m+1  \\
        \hline
        \hline
        22,8  & 0.338 & 0.316 & $\Gamma \to \Gamma$  & 3m \\ 
        22,10 & 0.317 & 0.326 & $X \to X$ & 3m$-$1  \\
        22,12 & 0.318 & 0.316 & $\Gamma \to \Gamma$ & 3m+1  \\
        \hline
    \end{tabular}
    \caption{DFT direct gap at $X$ ($\Delta _X$) and at $\Gamma$ ($\Delta _\Gamma$) in eV, the actual top valence and bottom conduction k-points and the family of the whole width $N$+$L$ for AGBNs of different size. Points labelled with a fraction are placed along the $\Gamma-X$ line with $X=\nicefrac{1}{2}$. }
    \label{tab:gap_AGBN_DFT}
\end{table}


In Table~\ref{tab:gap_AGBN_DFT} we group together the DFT direct gap at X ($\Delta_X$) and at $\Gamma$ ($\Delta_\Gamma$) for different $N$ in a range from $N$=2 to $N$=22. 
It actually comes out that the gap is often direct at $\Gamma$, but there are some exceptions.
Namely, in those structures where $N$+$L$=$3m-1$, DFT often predicts a direct gap at $X$, so in a symmetry point perpendicular to the ribbon axis. 
%
In some cases, the gap happens to be indirect between $X$ and $\Gamma$ or involving a point along the $\Gamma-X$ line.
Note, however, that the band dispersion perpendicular to the ribbons' axis is very weak and the maximum energy difference we observed between a gap at $X$ and at $\Gamma$ is around 60~meV. 
These energy differences are within the limits of precision of our DFT calculations and may be affected by the choice of some parameters and approximations like the exchange-correlation potential or the relaxation threshold.
We have not investigated further the topic because a fine characterisation of the $\Gamma-X$ dispersion is not relevant for the scope of our article.

\begin{figure*}
    \centering
    \includegraphics[width=\linewidth]{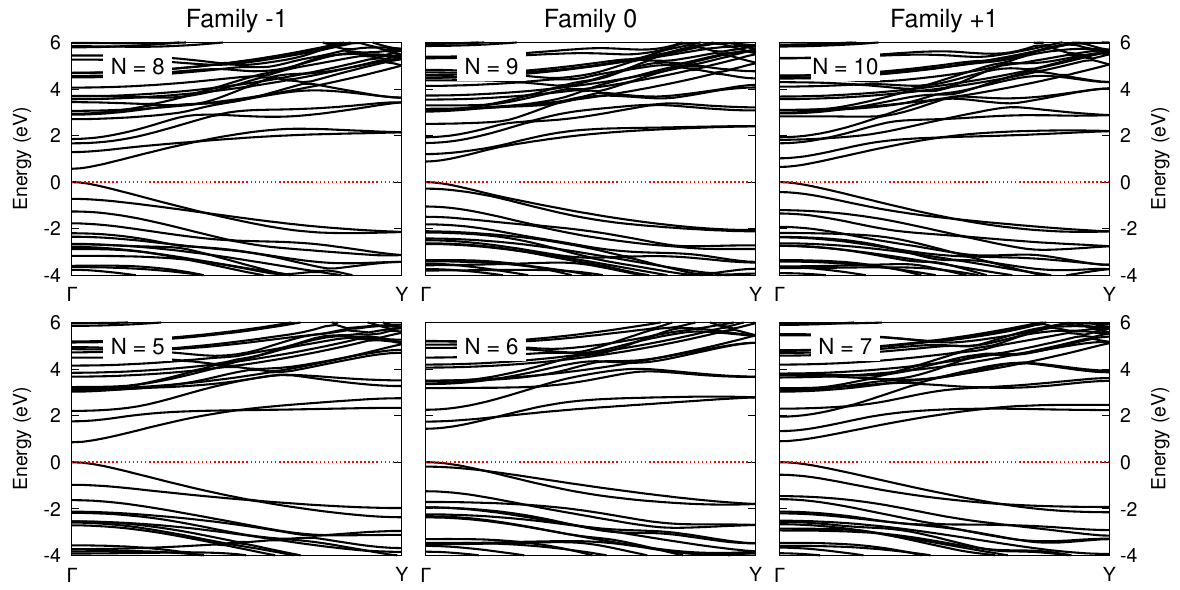}
    \caption{DFT band structures for systems with $N$=5 to $N$=10.}
    \label{fig:all_bands}
\end{figure*}



\bigskip

\section*{APPENDIX C: BAND STRUCTURES FOR N RANGING FROM 5 TO 10 }
In Figure~\ref{fig:all_bands} we report the DFT band structure along the $\Gamma$-$Y$ line for AGBNs with $N$ ranging from 5 to 10 and $L$ set either to 8 or 9 depending on the parity of $N$.
This allows us to show two representative band structures per ribbon family. 
Note that in the $N$=5 panel, we report the same band structure as in Figure~\ref{fig:dft-gw_gaps}.b.


\section*{APPENDIX D: LINK BETWEEN LADDER MODEL AND DFT BAND INDEXING}
In a heteroatomic ladder model, the unperturbed energies read:
\begin{eqnarray}
    E^0_{n\pm} &=& \bar{\epsilon} \pm \sqrt{\epsilon^2 + \tau_n^2} \,,\\
    \tau_n &=& t\left[ 1 + 2\cos\left( \theta_n \right)\right] \,,\\
    \theta_n &=& \pi n (N+1)^{-1} \,,
\end{eqnarray}
where $2\bar{\epsilon} = \epsilon_1 + \epsilon_2$ and $2\epsilon = \epsilon_1 - \epsilon_2$.

\begin{figure}
    \centering
    \includegraphics[height=0.90\textheight]{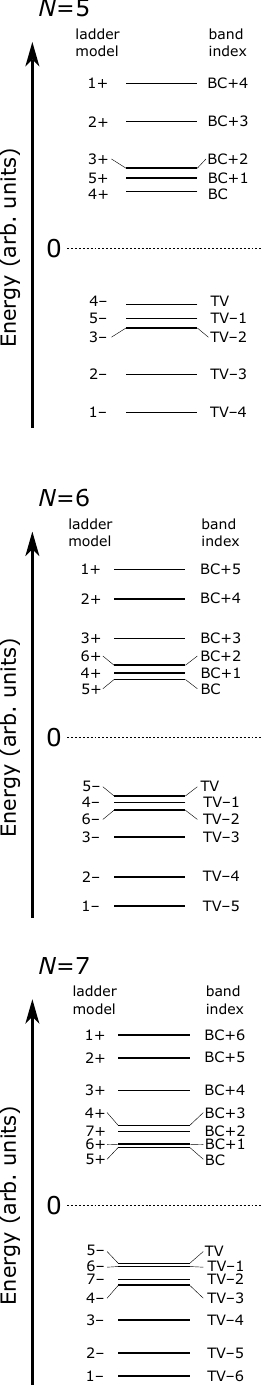}
    \caption{From top to bottom: Ladder model state indexes $n\pm$ and DFT band index in nanoribbons with $N$=5, 6 and 7. Here, for sake of simplicity, $\epsilon_1 = -\epsilon_2 = 2 t$.}
    \label{fig:ladder_indexes}
\end{figure}

The resulting energies of occupied ($-$) and empty ($+$) states are arranged symmetrically with respect to the average on-site energy $\bar{\epsilon}$, though the sequence is not necessarily monotonic with $n$. 
As a result, the link between the DFT band index and the ladder model state index is not straightforward. 
To illustrate this, we plot in Figure~\ref{fig:ladder_indexes} the sequence of the unperturbed ladder model states for $N=5$, $N=6$ and $N=7$ indicating both the $n\pm$ labelling of the ladder model (on the left of each level) and the corresponding band index in the corresponding DFT calculation (on the right). 
Note that the sequence does not depend on the actual values of $\epsilon_1$, $\epsilon_2$ and $t$.

\section*{APPENDIX E: ELECTRON LOCALIZATION FUNCTIONS}
\begin{figure}
    \centering
    \includegraphics[width=0.96\linewidth]{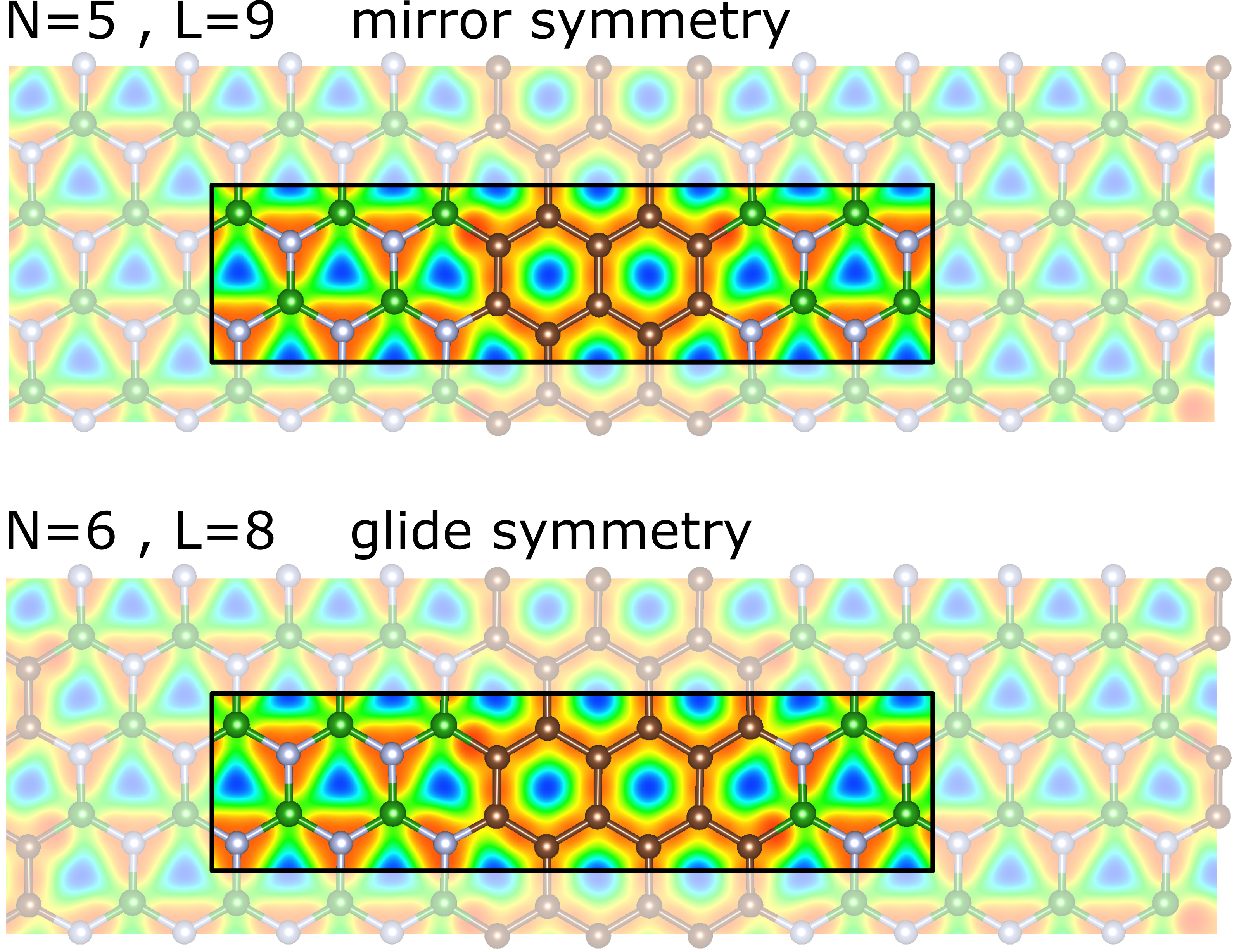}
    \caption{Ball-and-stick models reference systems shown in Figure~\ref{fig:ag5bn9_structure} superposed to the Electron Localization Function (ELF). Brown, green and grey spheres stand for C, B and N atoms. An unshaded rectangle with black borders draws the unitary cell in each system. ELF's color scale ranges from blue (minima) to red (maxima) in arbitrary units.}
    \label{fig:elf}
\end{figure}

In Figure~\ref{fig:elf} we report the electron localization function (ELF) of the $(N=5,L=9)$ and $(N=6,L=8)$ heterostructures of Figure~\ref{fig:ag5bn9_structure}.
The color gradient ranges from blue to red, the latter marking the maxima of the ELF. 
In the BN ribbons, electrons localize on the N atoms. 
Here, ELF maxima draw typical triangular shapes proper to strongly polarized sp$^2$ bonding. 
Conversely, in the Gr regions, electrons localize equally on the C atoms and C-C bonding, as expected in perfectly covalent sp$^2$ bonds.

\newpage


\bibliography{biblio}

\end{document}